\documentstyle[12pt,psfig]{ioplppt}

\begin{document}
\jl{3}

\newcommand{\chemical}[1]{{$\rm #1$}}
\renewcommand{\vec}{\bf}
\newcommand{\ti}[1]{{\tilde{\mathrm #1}}}
%
\title{Transitions between Phases with Equal Wave Numbers in a Double Ising Spin Model.
Application to Betaine Calcium Chloride Dihydrate}[Transitions between Phases with Equal Wave Numbers]
\author{Boris Neubert\dag, Michel Pleimling\ddag\ and Rolf Siems\dag}
\address{
\dag Theoretische Physik, 
Universit\"at des Saarlandes, 
Pf.\ 151150, 
D-66041 Saarbr\"ucken,
e-mail: neubert@lusi.uni-sb.de}
\address{
\ddag Theoretische Physik, 
Rheinisch-Westf\"{a}lische Technische Hochschule, 
Sommerfeldstra{\ss}e,  
D-52074 Aachen, e-mail: pleim@physik.rwth-aachen.de}
%
\begin{abstract}
A Double Ising Spin model for uniaxially structurally modulated
materials exhibits as a special feature phase
transitions between phases with equal wave numbers but
different pseudo spin configurations. 
The character of these `internal' transitions is investigated in mean field approximation,
with the mean field transfer matrix method, and in Monte Carlo simulations.
The structural changes at the transitions are characterized by 
different strengths of harmonics in a Fourier analysis of the spatial modulation.
A dielectric anomaly in the phase diagram of betaine calcium chloride dihydrate 
(BCCD) and seemingly contradictory structure analyses are explained.
\end{abstract}
\maketitle

\section{Introduction}

Materials exhibiting commensurately and incommensurately structurally 
modulated phases were extensively investigated during the last years. 
BCCD, \chemical{(CH_3)_3NCH_2COO\cdot CaCl_2\cdot H_2O}, is an outstanding 
example: it shows a rich variety of phases,
and exhibits many intricate phenomena in its phase diagram, such as
indications for accumulation points of structure branchings, and the so-called
dielectric $T_S$-anomaly \cite{Sch98}. The problem of the
modulation and its temperature-dependence still poses many open questions.
The plenitude of experimental results led to a continuing theoretical interest,
and many different models for their explanation have been
developed (for reviews see references \cite{Reviews,Neu98a}).\par
We recently used
the concept of symmetry-adapted local modes \cite{Tho71} to derive
the Double Ising Spin (DIS) model \cite{Ple94,Ple96} 
as a symmetry-based pseudo spin model with two 
two-component pseudo spin variables per crystallographic unit cell: the 
symmetry-breaking atomic displacements occuring at the transition from 
the high-temperature (high-symmetry) para phase to the modulated phases at lower temperatures
are expanded in terms of an appropriately chosen basis set, and 
the pseudo spin components $\tau$ and $\sigma$ are the signs of the relevant 
coordinates \cite{Neu98b,Neu98c}. In the case of materials 
with high-temperature phase space group $Pnma$ (e.g.\ BCCD), this approach yields the
Hamiltonian of the DIS model \cite{Ple94}:
\begin{eqnarray}
\fl
H = K \sum\limits_{ijk} \tau_{ijk} \tau_{ij(k+1)} + L \sum\limits_{ijk}
\sigma_{ijk} \sigma_{ij(k+1)} + \frac{M}{2} \sum\limits_{ijk} \left(
\sigma_{ijk} \tau_{ij(k+1)} - \tau_{ijk} \sigma_{ij(k+1)} \right)  \nonumber\\
\lo+ J \sum\limits_{ijk} \tau_{ijk} \left( \tau_{(i+1)jk} + \tau_{i(j+1)k}
\right) + J' \sum\limits_{ijk} \sigma_{ijk} \left( \sigma_{(i+1)jk} +
\sigma_{i(j+1)k} \right).
\label{equation:dis}
\end{eqnarray}
$i$, $j$, $k$ label the pseudo spin positions
along the $\bf a$, $\bf b$, $\bf c$-directions respectively. At $T = 0$,
the in-layer couplings $J<0,\,J' < 0$ lead to a ferro-ordering of the
layers perpendicular to ${\bf c}$. 
In ${\bf c}$-direction, frustrations and therefore modulations arise 
because of the antagonistic
effects of the symmetric nearest neighbour interactions $K$ and $L$
on the one hand and the antisymmetric interaction $M > 0$ on the other hand.
The stable structures
(pseudo spin profiles)
obtained from the statistical mechanics treatment of $H$ are characterized
by their wave numbers $q$ given in units of $2\pi/$(pseudo spin spacing).\par
At $T=0$, five stable phases are separated by multiphase lines in the $K/M$-$L/M$-plane.
Along these lines, an
infinity of different phases are degenerate. We restrict ourselves
to ferroelectric $\tau$-$\tau$-interactions, i.e.\ $K < 0$; the phase diagram
for $K > 0$ is obtained by exploiting the properties of $H$.
In the following, we will consider the multiphase lines (a) and (b) separating, in the
notation of reference \cite{Ple96}, phase V 
(the pseudo spin structure repeats itself after four
layers), from the ferroelectric phase I (all $\tau$- and $\sigma$-spins 
in ferro-order) and the mixed phase III ($\tau$-spins
in ferro-, $\sigma$-spins in antiferro-order along $\bf c$) respectively.\par
For low temperatures, all phases
degenerate at line (a) (with $K < 0$ and $L < 0$) 
and obeying some rules have a finite
stability range in the vicinity of this line \cite{Ple96}. -- Incidentally,
this also holds for the four-state chiral clock
model which can be derived from the DIS model for the case $K = L < 0$ 
and $J = J' < 0$ \cite{Ple98}.\par
In contrast to this,
in the vicinity of line (b) (with $K < 0$ and $L >0$) only four
different phases (characterized by the wave numbers $q = 0$, $\frac{1}{8}$,
$\frac{1}{6}$, and $\frac{1}{4}$) are stable at very low temperatures \cite{Ple96}.
Other phases
are created at higher temperatures by structure combination branching processes.\par
In the present paper, we report on a remarkable type of transitions
in the DIS model for the case $K < 0$ and $L > 0$: these `internal' transitions occuring
at certain temperatures $T_{\mbox{\small int}}$ are characterized by a structural change 
in the
pseudo spin arrangement which is not accompanied by a change of $q$ and -- in some cases --
not even by a change of the symmetries of the profile.\par
The paper is organized as follows. In section \ref{section:MFA}, the pseudo spin profiles
calculated in mean field approximation (MFA) are discussed. The results are substantiated in 
section \ref{section:Validation} by employing the mean field transfer matrix
method and Monte Carlo simulations, which prove that these transitions are no artefacts 
of MFA. In section \ref{section:BCCD},
the internal transition in the $q=\frac{1}{8}$-phase of the DIS model is used
to explain a dielectric anomaly in the fourfold phase of BCCD as well as
seemingly contradictory experimental structure analysis results. 
Section \ref{section:discussion} contains a summarizing discussion.\par
%

\section{Mean field investigation of the internal transitions}
\label{section:MFA}

The mean field approximation is often used for analysing the finite temperature
behaviour of pseudo spin models.
Starting point is the variational Hamiltonian
\begin{equation}
H_0 = - \sum\limits_{ijk} \underline{\eta}_k \cdot \underline{S}_{ijk}
\end{equation}
for systems with a modulation in $(001)$-direction,
with $\underline{S}_{ijk} = \left( \tau_{ijk}, \sigma_{ijk} \right)$.
The product $\underline{\eta}_k \cdot \underline{S}_{ijk}$
describes the interaction of the spin 
$\underline{S}_{ijk}$ (which in our case has two components) with a site-dependent 
mean field $\underline{\eta}_k$ depending on the averages of the neighbouring spins.
The mean values $t_\ell = \left< \tau_{ij\ell} \right>$ and
$s_\ell = \left< \sigma_{ij\ell} \right>$ of the pseudo spins $\tau$ and $\sigma$ in the
layer $\ell$ are determined by the MFA equilibrium equations
($ \beta = \frac{1}{k_B T}$):
\begin{eqnarray}
t_\ell & = & \tanh \left( \beta \eta_{1,\ell} \right), \nonumber \\
s_\ell & = & \tanh \left( \beta \eta_{2,\ell} \right),
\label{equation:mfaeq}
\end{eqnarray}
with the mean fields
\begin{eqnarray}
\eta_{1,\ell} & = & -  4 J t_\ell - K \left( t_{\ell+1} + t_{\ell-1} \right)
+ \frac{M}{2} \left( s_{\ell+1} - s_{\ell-1} \right),
\nonumber \\
\eta_{2,\ell} & = & -  4 J' s_\ell - L \left( s_{\ell+1} + s_{\ell-1} \right)
- \frac{M}{2} \left( t_{\ell+1} - t_{\ell-1} \right) \nonumber .
\end{eqnarray}
Figure \ref{figure:figure1} shows a typical mean field phase diagram in the plane
$\lambda = 0.1$, 
$j = j' = - 0.25$, where we have introduced the reduced interaction parameters
$\kappa = \frac{K}{M}, ~\lambda = \frac{L}{M}, ~j = \frac{J}{M},
~j' = \frac{J'}{M}$ and $\kappa_- = \frac{1}{2} ( \kappa - \lambda )$;
$\theta = \frac{k_B T}{M}$ denotes the reduced temperature.
The shaded areas contain incommensurate and higher order commensurate phases.
The different commensurate phases are characterized by their wave numbers.
Para phase, ferro phase and modulated phases meet at the
Lifshitz point (LP). At $\theta = 0$ the considered cut through the 
temperature-interaction phase diagram intersects the multiphase line
at the multiphase point MP.\par
The existence of accumulation points of structure branchings
in the case $K < 0$ and $L > 0$ was verified  \cite{Ple94} by
formulating the mean field equilibrium equations (\ref{equation:mfaeq}) 
as a four-dimensional mapping and
using a fixed-point technique analogous to that employed for the ANNNI model
\cite{FPZit}. Also calculated were the self- and the interaction-energies of
discommensurations.\par
As a consequence of the signs of the parameters $K$, $L$ and $M$ (i.e.\ $K < 0$, $L > 0$, and
$M > 0$), all phases degenerate at the multiphase line (b) consist  
of bands of two types of two-layer sequences 
viz.\ (\raisebox{-1ex}{$\stackrel{\textstyle+}{-}$}
\raisebox{-1ex}{$\stackrel{\textstyle+}{+}$}) and
(\raisebox{-1ex}{$\stackrel{\textstyle-}{+}$}
\raisebox{-1ex}{$\stackrel{\textstyle-}{-}$}), where the upper (lower) symbols represent
the signs of the $\tau$- ($\sigma$-) spin.
We introduce a bracketed notation 
$\left< A_1 \ldots A_k\right>$
for the different phases: for example, $\left<23\right>$ denotes a repetitive pseudo 
spin structure consisting of a band with two (\raisebox{-1ex}{$\stackrel{\textstyle+}{-}$}
\raisebox{-1ex}{$\stackrel{\textstyle+}{+}$})
sequences followed by a band with three
(\raisebox{-1ex}{$\stackrel{\textstyle-}{+}$}
\raisebox{-1ex}{$\stackrel{\textstyle-}{-}$}) sequences.
The wave number for the above phase is 
\begin{displaymath}
q = \frac{k}{4 ~ \sum\limits_{i = 1}^k ~ A_i}.
\end{displaymath}
The equilibrium pseudo spin profiles for configurations with a fixed wave number $q$
at a given temperature are obtained by finding the self-consistent solutions of
the equations (\ref{equation:mfaeq}). This is done by starting from different initial
spin configurations with the same modulation period and iterating. The respective free
energies are calculated and the solution with the
lowest free energy describes the stable spin configuration. 
Keeping $\lambda$, $j$, and $j'$ fixed, the reduced parameters
$\theta$ and $\kappa_-$
are then varied in such a way
that the resulting line $\theta\left(\kappa_-\right)$ always lies wholly in the 
stability region of the phase under investigation.
Figure \ref{figure:figure2} shows the obtained 
spin profiles for the phases $\left< 12 \right>$ (wave number $q = \frac{1}{6}$)
and $\left< 2 \right>$ ($q = \frac{1}{8}$) with $\lambda = 0.05$ and
$j = j' = -1$. For both wave numbers there is a discontinuous internal
phase transition. At higher temperatures, the $q = \frac{1}{6}$-phase
exhibits also a continuous transition to a modification with two disordered layers.\par
The obtained spin profiles for the $\left< 2 \right>$-phase (for layers 
1 to 8, i.e.\ for one modulation period) exhibit symmetries given by
\begin{itemize}
\item[(a)] at low temperatures:
\begin{displaymath}
\begin{array}{lrrrrrrrr}
t:~~~ &t_1& t_2&t_2&t_1&-t_1& -t_2& -t_2& -t_1\\ 
s:~~~ &- s_1&s_2& -s_2& s_1& s_1& -s_2& s_2& -s_1
\end{array}
\end{displaymath}
\item[(b)] at high temperatures:
\begin{displaymath}
\begin{array}{lrrrrrrrr}
t:~~~ & t_1& t_2& t_2& t_1& -t_1& -t_2& -t_2& -t_1 \\
s:~~~ & - s_1& -s_2& s_2& s_1& s_1& s_2& -s_2& -s_1. 
\end{array}
\end{displaymath}
\end{itemize}
At a reduced temperature $\theta_{\text{\small int}}$ the $\sigma$-spins averages 
change signs in layers 2, 3, 6 and 7, whereas 
the symmetry elements of the spin profiles are left unchanged. Furthermore, also the
$\tau$-spins (in all layers) show discontinuities at $\theta_{\text{\small int}}$.\par
For the $\left< 12 \right>$-phase, the spin profiles in the vicinity of the discontinuous
transition are:
\begin{itemize}
\item[(a)] at low temperatures:
\begin{displaymath}
\begin{array}{lrrrrrr}
t~~~: & t_1& t_1& -t_2& -t_3& -t_3& -t_2 \\
s~~~: & - s_1& s_1& s_2& -s_3& s_3& -s_2 
\end{array}
\end{displaymath}
\item[(b)] at high temperatures:
\begin{displaymath}
\begin{array}{lrrrrrr}
t~~~: & t_1& t_2& t_3& -t_1& -t_2& -t_3 \\
s~~~: & - s_1& s_2& s_3& s_1& -s_2& -s_3. 
\end{array}
\end{displaymath}
\end{itemize}
In this case the $\tau$-spins in layer 3 (and the $\sigma$-spins in layers 4 and 5) 
change signs, leading to a
change of the symmetries of the profile.\par
In general not all combinations of the signs of the spins were taken
as possible starting points for solving the equilibrium equations self-consistently.
However, we substantiated our results for selected values of the parameters by
considering all $2^{11}$ (phase $\left< 12 \right>$) or $2^{15}$ (phase $\left< 2 \right>$)
possible sign combinations (changing the signs of all spins leaves the free 
energy unchanged).\par
In order to characterize the different modifications, a Fourier analysis of the spatial 
modulation  of $t_\ell$ and $s_\ell$ was performed. In the following, the results 
for the $q = \frac{1}{8}$-phase are discussed, as they will be needed for the application 
to BCCD in section
\ref{section:BCCD}.
For reduced temperatures slightly below ($\theta= 2.967$) and above
($\theta= 2.993$) the reduced temperature $\theta_{\text{\small int}}$ of the internal transition,
the pseudo spin averages (profiles) for one period of the spatial modulation 
are shown in
figure \ref{figure:Profile} together with their harmonics.
Due to the symmetries in the pseudo spin profiles, the Fourier components of $t_\ell$
($s_\ell$) are all odd (even) about $\ell=\frac{1}{2}$, and only odd orders in sine (cosine) 
occur in the Fourier series:
\begin{eqnarray}
\label{equation:harmt}
t_\ell &= & a_1 \sin(\frac{\pi}{4} [\ell-\frac{1}{2}]) + a_3 \sin(\frac{3\pi}{4} [\ell-\frac{1}{2}])\\
\label{equation:harms}
s_\ell &= & b_1 \cos(\frac{\pi}{4} [\ell-\frac{1}{2}]) + b_3 \cos(\frac{3\pi}{4} [\ell-\frac{1}{2}]).
\end{eqnarray}
Upon cooling (see figure \ref{figure:Harmonics}),
the relative strength of the third order harmonic with respect to the first order harmonic
increases smoothly from zero near the transition to the para phase to about one third 
near $\theta_{\text{\small int}}$ (both for the $t$- and $s$-profiles).
When dropping below $\theta_{\text{\small int}}$, 
only a slight increase in $a_1$ and $a_3$ but remarkable changes in $b_1$ and $b_3$ occur:
$b_3$ jumps
from $0.391$ at $\theta= 2.993$ to $-1.089$ at $\theta= 2.967$ while $\left| b_1 \right|$ decreases from
$1.136$ to $0.536$. Below $\theta_{\text{\small int}}$, the $s$-profile is governed by the strong third order harmonic
($|b_3/b_1| > 2.03$),
which continuously increases to the ground state value $|b_3|= 1.307$, whereas the first harmonic keeps 
its nearly constant value $\left| b_1 \right|= 0.54$.\par 
In MFA, these discontinuous internal phase transitions were obtained 
for all investigated phases of the DIS model, execpting only
the $\left< 1 \right>$- and
the $\left< \infty \right>$- (ferro) phases. They occur
for all values of $\lambda$ and $\kappa$ as long as
$\lambda > 0$ and $\kappa < 0$, as well as for all
in-layer couplings $j$ and $j'$ (even for very large values).

\section{Verification by mean field transfer matrix and Monte Carlo methods}
\label{section:Validation}

In the previous section, the DIS model was treated in the mean field approximation.
The question arises whether the observed internal transitions are an artefact
of this approximation or if they are also present if one goes beyond mean field theory.\par
This problem should indeed be considered: in MFA,
the Axial Next Nearest Neighbour Ising (ANNNI)
model  \cite{Sel88} with weak in-layer couplings exhibits, for example, phase transitions
from the low temperature phases to high temperature asymmetric and partially
disordered phases \cite{Yok91}. 
Nakanishi  \cite{Nak92} could show that these phases are MFA artefacts by demonstrating that
they do not occur if the mean field transfer matrix (MFTM) method is used, in which
the interactions in direction of the modulations are treated exactly.
The same conclusion was drawn
from Monte Carlo simulations  \cite{Rot93}.\par
In the MFTM method, the variational Hamiltonian for the DIS model is
\begin{eqnarray}
\label{equation:HMFTM}
H_0 &=& K \sum_{i j k} \tau_{i j k} \tau_{i j (k+1)} + L \sum_{i j k} \sigma_{i j k} \sigma_{i j (k+1)}  \nonumber \\
&&+ \frac{M}{2} \sum_{i j k} (\sigma_{i j k} \tau_{i j (k+1) } - \sigma_{i j (k+1)} \tau_{i j k})
- \sum\limits_{ijk} \underline{\eta}_k \cdot \underline{S}_{ijk}.
\end{eqnarray}
In the MFA variational Hamiltonian (\ref{equation:dis})  all
interactions are described by mean fields. In contrast, in equation (\ref{equation:HMFTM})
site-dependent mean fields are only introduced
for the interactions perpendicular to the direction of modulation. As the
two-component mean fields $\underline{\eta}_k$ do not depend on the
indices $i$ and $j$ the problem is reduced to one dimension
and the indices $i$ and $j$ will be omitted.\par
Using the Bogoliubov variational principle,
the MFTM free energy is given by
\begin{equation}
F = 2 J \sum\limits_\ell t_\ell^2 + 2 J' \sum\limits_\ell s_\ell^2 +
\sum\limits_\ell \eta_{1,\ell} \, t_\ell + \sum\limits_\ell \eta_{2,\ell} s_\ell
- k_B T \ln Z
\label{equation:mftmfree}
\end{equation}
with
\begin{equation}
\fl Z = \mbox{Tr} \, \exp \left( \ti{K} \sum\limits_\ell \tau_\ell \, \tau_{\ell+1} +
\ti{L} \sum\limits_\ell \sigma_\ell \, \sigma_{\ell+1} + \frac{\ti{M}}{2}
\sum\limits_\ell \left( \tau_{\ell+1} \, \sigma_\ell - \sigma_{\ell+1} \, \tau_\ell \right)
+ \sum\limits_\ell \left( h_{1,\ell} \, \tau_\ell + h_{2,\ell} \, \sigma_\ell \right) \right)
\label{equation:mftmpart}
\end{equation}
and $\ti{K} = \beta K$,~ $\ti{L} = \beta L$,~ $\ti{M} =
\beta M$ and $h_{1,\ell} = \beta \eta_{1,\ell}$,~ $h_{2,\ell} = \beta \eta_{2,\ell}$.
In order to calculate the partition function $Z$ we introduce the
transfer matrix
\begin{displaymath}
\fl
T_\ell = \left( \begin{array}{llll}
e^{h_{1,\ell}+h_{2,\ell}} & 0 & 0 & 0 \\
0 & e^{h_{1,\ell}-h_{2,\ell}} & 0 & 0 \\
0 & 0 & e^{-h_{1,\ell}+h_{2,\ell}} & 0 \\
0 & 0 & 0 & e^{-h_{1,\ell}-h_{2,\ell}}
\end{array} \right) \, \left( \begin{array}{llll}
e^{\ti{K}+\ti{L}} & e^{\ti{K}-\ti{L}+\ti{M}} & e^{-\ti{K}+\ti{L}-\ti{M}} &
e^{-\ti{K}-\ti{L}} \\
e^{\ti{K}-\ti{L}-\ti{M}} & e^{\ti{K}+\ti{L}} & e^{-\ti{K}-\ti{L}} &
e^{-\ti{K}+\ti{L}+\ti{M}} \\
e^{-\ti{K}+\ti{L}+\ti{M}} & e^{-\ti{K}-\ti{L}} & e^{\ti{K}+\ti{L}} &
e^{\ti{K}-\ti{L}-\ti{M}} \\
e^{-\ti{K}-\ti{L}} & e^{-\ti{K}+\ti{L}-\ti{M}} & e^{\ti{K}-\ti{L}+\ti{M}} &
e^{\ti{K}+\ti{L}}
\end{array} \right).
\end{displaymath}
For a commensurate phase with period $p$ the mean fields repeat
themselves after $p$ layers. The partition function (\ref{equation:mftmpart}) is then given by
$Z = \alpha_{max}^{N/p}$
where $\alpha_{max}$ is the maximum eigenvalue of the matrix product
$\prod\limits_{\ell=1}^p T_\ell$ and $N$ is the number of layers in the crystal.\par
The mean values $t_\ell$ and $s_\ell$ are calculated by
\begin{eqnarray}
t_\ell &=& \frac{\partial \ln \alpha_{max}}{\partial h_\ell^1}
= \frac{1}{\alpha_{max}} \left< \alpha_{max} \left| T_1 ... T_{\ell-1}
\mu^t
T_\ell ... T_p \right| \alpha_{max} \right> \nonumber \\
s_\ell &=& \frac{\partial \ln \alpha_{max}}{\partial h_\ell^2}
= \frac{1}{\alpha_{max}} \left< \alpha_{max} \left| T_1 ... T_{\ell-1}
\mu^s T_\ell ... T_p \right| \alpha_{max} \right>.
\label{equation:mftmeq}
\end{eqnarray}
$\left< \alpha_{max} \left| \right. \right. $ and
$ \left. \left. \right| \alpha_{max} \right> $ are the
left and the right eigenvectors corresponding to the eigenvalue
$\alpha_{max}$, whereas the matrices $\mu^t$ and $\mu^s$ are given by

\begin{displaymath}
\mu^t = \left( \begin{array}{llll}
1 & ~~~0 & ~~~~~0 & ~~~~0 \\
0 & ~~~1 & ~~~~~0 & ~~~~0 \\
0 & ~~~0 & ~~ -1 & ~~~~0 \\
0 & ~~~0 & ~~~~~0 & ~~ -1
\end{array} \right)
~~~~~\mbox{and}~~~~~
\mu^s = \left( \begin{array}{llll}
1 & ~~~~0 & ~~~0 & ~~~~0 \\
0 & ~~ -1 & ~~~0 & ~~~~0 \\
0 & ~~~~0 & ~~ ~1 & ~~~~0 \\
0 & ~~~~0 & ~~~0 & ~~ -1
\end{array} \right) .
\end{displaymath}
The mean fields
\begin{equation}
\eta_{1,\ell} = - 4 J t_\ell \hspace*{1cm} \mbox{and} \hspace*{1cm} \eta_{2,\ell} = 
-4 J' s_\ell
\label{equation:mftmfield}
\end{equation}
result from the minimization of the free energy (\ref{equation:mftmfree}).
Equations (\ref{equation:mftmeq}) and (\ref{equation:mftmfield})
are to be solved self-consistently.\par
As an example, we will discuss the application of this method to the $q = \frac{1}{8}$
phase. Calculations in mean field approximation (see previous section) yield a first order
phase transition at $\theta_{\text{\small int}}$ between a low- and a high-temperature modification
(figure \ref{figure:figure2}). These
two spin profiles are used as starting points for the self-consistent solution of the
MFTM equations (\ref{equation:mftmeq}) and (\ref{equation:mftmfield}).\par
Figure \ref{figure:figure5} shows the resulting free energies obtained by varying the
model parameters in the same manner as in the previous section,
i.e.\ one proceeds along the same $\theta(\kappa_-)$-curve as for the MFA calculations.
It is clearly seen that the free energy curves of the two modifications intersect at
$\theta_{\text{\small int}} = 2.91$, yielding a first-order phase transition. 
As expected, the temperature
$\theta_{\text{\small int}}$ is slightly shifted to lower temperatures as compared to the 
MFA result.\par
In the MFTM method some pseudo spin interactions, though not those in modulation
direction (which should be the most important ones),
are still replaced by an interaction of the pseudo spins with a mean field. 
To make sure that this approximation is not responsible 
for the occurrence of internal phase transitions,  Monte Carlo simulations were performed
for phases with a short modulation length. No attempt was
made to determine the global phase diagram.\par
Systems of sizes $L \times L \times M$ were simulated, with $L$ ranging from 10 to 40.
The number of layers, $M$, is given by the modulation length of the phase under
investigation, e.g.\ $M = 6$ for the $q = \frac{1}{6}$-phase. This is done in order
to avoid the development of fluctuations resulting from (meta)stable phases with
a larger modulation length. The standard one--spin--flip Metropolis algorithm is
used.\par
Figure \ref{figure:figure6} shows the calculated specific heat as a function of the reduced 
temperature for the $q = \frac{1}{6}$-phase, for interactions given by
$\lambda = 0.1$, $\kappa = -0.57$, and $j = j' = -0.5$ and a size of $30 \times 30
\times 6$. There are two different peaks. The peak at higher temperatures corresponds
to a second order phase transition to the disordered phase. In the present context, the 
narrow peak is the interesting one. It corresponds to 
the internal first order phase transition. The resulting spin profiles for temperatures
below and above the peak position compare
favorably with the profiles obtained in MFA. In the inset showing the calculated energy
the discontinuous character of the internal phase transition is clearly
visible.\par

\section{Application to BCCD}
\label{section:BCCD}

Due to the pseudo symmetry of half a lattice constant along ${\vec c}$, BCCD can be considered 
to be built up of successive layers of half cells perpendicular to ${\vec c}$, which we
label by the subscript $\ell$.
Applying the method described in reference \cite{Neu98b} to BCCD, the symmetry-breaking atomic
displacements occuring below the transition from the unmodulated high-temperature phase to the 
modulated phases are expanded in terms of a
localized symmetry-adapted basis set; the respective coordinates (mode amplitudes) 
are projected onto two-valued
pseudo spin variables leading to symmetry-based pseudo spin models. 
Every basis vector (symmetry-adapted local mode, SALM)
describes a collective displacement of all atoms in a half cell.
One obtains two relevant SALMs for the front and two for the back half cell. Adequate
superpositions of these describe structural modulations transforming according 
to the irreducible 
representations $\Lambda_2$ and $\Lambda_3$ of the group of the wave vector ${\vec q}= q {\vec c}^*$
(antisymmetric with respect to the mirror plane).\par
The pseudo spins $\tau$ and $\sigma$ represent the signs of the amplitudes of
the two relevant SALMs. For convenience, we shall henceforth call the latter 
,$\tau$-SALM' and ,$\sigma$-SALM'.
Previously \cite{Neu98a,Neu98c}, we deduced the displacements corresponding to the $\tau$- and
the $\sigma$-SALMs for all BCCD atoms from experimental data \cite{Ezp92}. 
For a complete visualization of these SALM displacements
it is sufficient to follow the $y$-displacements of the nitrogen atoms.
Consider the two nitrogen atoms 1 and 2 associated to a half cell. The major part of 
their $y$-displacements is provided by the $\tau$-SALM contributions (equal for 1 and 2).
The $\sigma$-SALM contributions (of equal sizes but opposite signs for 1 and 2) 
account for the small relative $y$-displacement of 1 and 2.\par
Taking the transformation properties of the SALMs into account,  
the form (\ref{equation:dis}) of the pseudo spin Hamiltonian is obtained
as well as expressions for the
spontaneous polarization in terms of the pseudo spins \cite{Neu98a,Neu98c}.
\par
Experimental investigations of the pressure-temperature ($p$-$T$) phase 
diagram of BCCD \cite{Sch96,Mai96,Mai97} 
revealed a dielectric anomaly in the modulated phase with 
wave vector ${\vec k}=\frac{1}{4} {\vec c}^*$.
It occurs along a line $T_S(p)$,
which runs almost
parallel to the boundary of the para phase and to the lines
of constant spontaneous polarization $P_{S,x}$. At ambient pressure, $T_S$ 
practically coincides with
the upper boundary of the fourfold phase. 
The $T_S$-anomaly exhibits the characteristics of
a first order phase transition; it leaves the wave number unchanged. 
One proposal for the explanation of the $T_S$-anomaly was to interpret it
as a freezing of the soliton lattice \cite{Mai97}, i.e.\ to assume that 
the soliton lattice is free to move above $T_S$ and is pinned below.\par
Since the wave number $k$ in BCCD is measured in units of the reciprocal lattice constant,
whereas the wave number $q$
in the DIS model is given in units of the reciprocal layer spacing, the $k=\frac{1}{4}$-phase in BCCD
corresponds to the $q=\frac{1}{8}$-phase in the DIS formalism.
For the latter the occurence of an internal transition was explicitely shown in sections
\ref{section:MFA} and \ref{section:Validation} above.
It suggests itself to interpret the $T_S$-anomaly as a manifestation
of the internal transition
observed in the DIS model. This hypothesis is supported by the following results:\par
In the same way as the 
$T_S$-line in BCCD, the $T_{\mbox{\small int}}$-line runs parallel to the boundary 
of the high-temperature phase and 
parallel to the lines of constant spontaneous polarization \cite{Neu98a} 
$P_{S,x}= \sum_\ell (-1)^\ell [P^1_x t_\ell t_{\ell+1} +  P^2_x s_\ell s_{\ell+1} + P^3_x s_\ell t_\ell +P^4_x \left( t_\ell s_{\ell+1} - s_\ell t_{\ell+1} \right)]$.\par
There is an interesting connection with experimental studies on the structure 
of the fourfold phase in BCCD:\par 
The displacement of atom $\mu$ in cell $\vec n$ can be written as (e.g.\ reference \cite{Her98})
\begin{equation}
\label{equation:AMF}
{\vec u}^\mu_{\vec n}= \sum_{m\geq 1} \left[ {\vec e}^{1\mu}_m \cos(2\pi m {\vec k} \cdot ({\vec r}_{\vec n}+{\vec \rho}^\mu)) + {\vec e}^{2\mu}_m \sin(2\pi m {\vec k} \cdot ({\vec r}_{\vec n}+{\vec \rho}^\mu)) \right],
 \end{equation}
where the zeroeth order harmonics are absorbed in the average positions ${\vec \rho}^\mu$ of $\mu$
with respect to the origin of the cell. There are only odd harmonics 
due to the symmetry of the $k=\frac{1}{4}$-phase in BCCD.
A comparison of the symmetries of the eigenvectors 
${\vec e}^{1\mu}_m$ and ${\vec e}^{2\mu}_m$ for odd $m$ and of the
$\tau$-SALM and the $\sigma$-SALM shows that the $\sin$ and $\cos$ terms in
(\ref{equation:AMF}) correspond to the $t$- and $s$-profile respectively \cite{Neu98c}.\par
Recent neutron scattering studies \cite{Her98} of the fourfold phase in BCCD at $100 K$ and ambient pressure
(see figure \ref{figure:Hypo}a),
i.e.\ below $T_S\approx 115K$, revealed a strong third harmonic in the $\cos$ terms in (\ref{equation:AMF}).
This is in perfect agreement with the structure of the
low temperature modification in the DIS model as derived in 
sections \ref{section:MFA} and \ref{section:Validation}.
X-ray techniques \cite{Ezp92} showed a more sinusoidal
modulation ($T=90K$, ambient pressure), which can be well described 
as the high temperature modification
in the DIS model, i.e.\ the one that is expected to be stable above $T_S$.\par
These seemingly contradictory experimental results can be explained if the influence of point
defects on the phase diagram and recent experiments on intentionally X-ray-irradiated
samples are taken into
consideration. Exposure to X-rays -- even to doses used normally in 
diffractometry \cite{Kia95,Mai96} -- produces radiation damage, which is sufficient to 
cause an observable lattice expansion. The effect is comparable to doping 
with bromine ions. The resulting (positive) plastic strains shift the boundaries
in the $p$-$T$-phase diagram to higher pressures/lower temperatures.
It was experimentally observed, that this shift is stronger for the $T_S$-line than
for the other phase boundaries \cite{Mai96,Mai97}.
This leads to the situation described schematically in figure \ref{figure:Hypo}b:
the irradiation defects induced by the X-ray measurements
produce different shifts of the $T_S$-line and the other boundaries 
and thus open up a window for the observation of the 
high temperature modification of the $k=\frac{1}{4}$-phase at ambient pressure.\par
For a more quantitative comparison of the structures 
predicted by the symmetry-based DIS model with the atomic displacements
observed either by X-ray or by neutron measurements, we derived the 
amplitudes $Q^\tau_\ell$ and $Q^\sigma_\ell$  
in a decomposition of the structural modulation in the 
fourfold phase of BCCD in terms of the
$\tau$- and $\sigma$-SALMs (see figure \ref{figure:BCCDProfile}) from
the experimental data given in references \cite{Her98,Ezp92}. 
In the employed procedure \cite{Neu98a}, $Q^\tau_\ell$ and $Q^\sigma_\ell$ are proportional 
to $t_\ell$ and $s_\ell$ in linear approximation.
$Q^\tau_\ell$ and $Q^\sigma_\ell$ can be written in terms of first and third
order harmonics as in equations\ (\ref{equation:harmt}) and (\ref{equation:harms}).
We obtained $a_3/a_1= 0.15$, $b_3/b_1= -0.08$ from the X-ray data \cite{Ezp92} and
$a_3/a_1= 0.27$, $b_3/b_1= 2.35$ from the neutron results \cite{Her98}.
Similar values can be
found in the DIS model for temperatures not too far above (e.g.\ $\theta= 4.25$: 
$a_3/a_1= 0.18$, $b_3/b_1= -0.07$) or below $\theta_{\text{\small int}}$ (e.g.\ $\theta= 2.76$: 
$a_3/a_1= 0.39$, $b_3/b_1= 2.13$) respectively. It is noteworthy that even with
the parameters ($\lambda= 0.05$, $j=j'=-1.0$) which were rather arbitrarily chosen in
this comparison and which probably do not represent the optimal choice
(e.g.\ by the methods discussed in ref.\ \cite{Neu98a}) for BCCD,
we find a good agreement between theoretical prediction
and experimental observation.\par
%
\section{Conclusions}
\label{section:discussion}
%
Recent theoretical developments in the field of
modulated systems allow, in a physically well
motivated way, to bridge the gap between pseudo spin model-parameters and
-properties on the one hand, and experimental control parameters and physical
characteristics of real systems on the other.\par
An efficient pseudo spin model, which is well suited as a basis
both for largely analytical calculations and for numerical simulations, is the DIS
model. It can advantageously be used for the theoretical description of a large class
of modulated materials. Since the model variables are explicitely related to local
properties of the discrete crystal lattice, it allows explicit predictions e.g.\ of
structures, symmetries, spontaneous polarizations, phase diagrams, and orders of
transitions of modulated phases in real systems.\par
A special feature of the DIS model is that it predicts phase transitions not only
between modulated structures with different wave numbers, but also between 
modulations with the same wave number but different pseudo spin configurations
(`internal' phase transitions). The respective phases differ significantly
in the amplitudes of the harmonics in a Fourier expansion of the pseudo spin profiles 
and in symmetry. Since remotely similar 
theoretical results derived for the ANNNI model were later shown to be artefacts
of the employed mean field approximation, we verified the validity of our MFA results
both by means of the mean field transfer matrix method and by Monte Carlo simulations.\par
One indication of the suitability of the symmetry-based DIS model for the
description even of detailed characteristics of complex modulated real systems is the
proposed explanation of the dielectric anomaly $T_S$ observed in the 
fourfold phase of BCCD: this anomaly and its characteristic properties can be well
interpreted as a realization of the internal phase transitions of the DIS model.\par
Furthermore, an application of the general procedure for translating model properties
to characteristics of real, experimentally investigated systems allows a tentative
explanation of discrepancies between experimental structure determinations of
the fourfold phase of BCCD by neutrons on the one hand and by X-rays on the other.
Taking the influence of X-ray induced defects (via the strain field) into account,
we propose a mechanism which allows the high temperature modification of the
fourfold phase (which exists in ideal crystals only at elevated pressure) to
become stable already at ambient pressure and, thus, be visible in X-ray measurements.
The two experimentally determined structures of the fourfold phase agree fairly well 
with our calculated data.
\clearpage
%
%
%
%
%
\ack
Stimulating discussions with Prof.\ Dr.\ J.M. P\'{e}rez-Mato, Bilbao, and financial
support from the Deutsche Forschungsgemeinschaft (research grant Si 358/2) are
gratefully acknowledged.
%


%
%
\Figures
\begin{figure}
\caption{Reduced temperature-interaction phase diagram of the DIS model with $\lambda = 0.1$
and $j=j'=-0.25$ (see text). Shown are the paraelectric, the ferroelectric, and
a few commensurate phases. Higher order commensurate and incommensurate phases
are contained in the hatched areas. LP is the Lifshitz point, MP the multiphase
point. $\theta$ is the reduced temperature $\frac{k_BT}{M}$, whereas $\kappa_-$
is given by $\kappa_- = \frac{1}{2} \left( \kappa - \lambda \right)$.
The phases are characterized by quotients $q=m/n$ giving the modulation wave numbers
in units of $2\pi/$(pseudo spin spacing).
\label{figure:figure1}}
\end{figure}
\begin{figure}
\caption{Pseudo spin profiles for (a) 
the $\left< 12 \right>$-phase ($q=\frac{1}{6}$) and (b) the $\left<2 \right>$-phase 
($q=\frac{1}{8}$) 
as functions of the reduced temperature $\theta$ with
$\lambda = 0.05$ and $j=j'=-1$. Shown are
the mean values $t_\ell$ (solid lines) and $s_\ell$ (dashed lines),
for layers $\ell$ from 1
to the modulation length, i.e.\ from 1 to 6 for case (a) and from 1 to 8 
for case (b).   \label{figure:figure2}}
\end{figure}
\begin{figure}
\caption{Pseudo spin profiles $t_\ell$ (upper part) and $s_\ell$ (lower part) 
of the $q=\frac{1}{8}$-phase for reduced temperatures
slightly below ($\theta= k_BT/M= 2.967$, left hand side) and slightly
above ($\theta= 2.993$, right hand side) the 
internal transition $\theta_{int}$. 
Only first (solid lines) and third (dashed lines) order harmonics in sine (cosine) 
are present in the Fourier expansion of $t_\ell$ ($s_\ell$).
\label{figure:Profile}}
\end{figure}
\begin{figure}
\caption{Amplitudes of the first and third order harmonics in the pseudo spin
profiles $t_\ell$ and $s_\ell$ of the $q=\frac{1}{8}$-phase as functions of the reduced 
temperature $\theta= k_BT/M$ [see equations (4) and (5)].
The internal phase transition at $\theta_{int}$ separates
the low temperature modification with 
dominant third harmonic in $s$ from the much less anharmonic high temperature modification.
\label{figure:Harmonics}}
\end{figure}
\begin{figure}
\caption{MFTM values for the reduced free energies $F/M$
of the low- (dashed line) and the high-temperature (solid line)
modifications of the $q = \frac{1}{8}$ phase as function of the
reduced temperature $\theta$. Fixed parameters are $\lambda = 0.05$ and $j=j'=-1$. 
Solid line: low temperature modification; dashed line: high temperature modification.
For better clarity, $F/M-\theta+4$ is plotted.\label{figure:figure5}}
\end{figure}
\begin{figure}
\caption{Specific heat of the $q = \frac{1}{6}$-phase obtained from
Monte Carlo simulations of DIS systems with $30 \times 30 \times 6$ spin sites.
$\lambda = 0.1$, $\kappa = -0.57$, and $j=j'=-0.5$. The inset shows
the calculated reduced energy as a function of $\theta$.\label{figure:figure6}}
\end{figure}
\begin{figure}
\caption{ $p$-$T$ phase diagram of BCCD near the $k=1/4$-phase. 
Solid lines: boundaries of the fourfold phase; dot-dashed lines: $T_S$-anomaly.
$\bullet$: neutron scattering measurements [19];
$\ast$: X-ray diffractometry [15].
a) Undisturbed sample (data from reference [18]). 
At ambient pressure, $T_S$ coincides with the upper phase boundary. 
b) Sample for which X-ray induced defects are (schematically) taken into account. 
The temperature decrease is larger for $T_S$ than for the phase boundaries. 
X-ray measurements at $\ast$ should establish the structure above $T_S$
even at ambient pressure.\label{figure:Hypo}}
\end{figure}
\begin{figure}
\caption{Fourfold phase of BCCD: amplitudes $Q^\tau_\ell$ and $Q^\sigma_\ell$ of the 
$\tau$- and $\sigma$-SALMs as determined by the corresponding decompositions of
neutron [19] (left hand side) and X-ray [15] (right hand side) data.
In a Fourier series expansion only first and third (solid and dashed lines 
respectively) order harmonics in sine (cosine) are present in the
spatial modulation of $Q^\tau_\ell$ ($Q^\sigma_\ell$). The neutron scattering 
results [19] show a much stronger third harmonic in the $\sigma$-SALM-profile 
than the X-ray diffraction measurements [15].
\label{figure:BCCDProfile}}.
\end{figure}
%
%

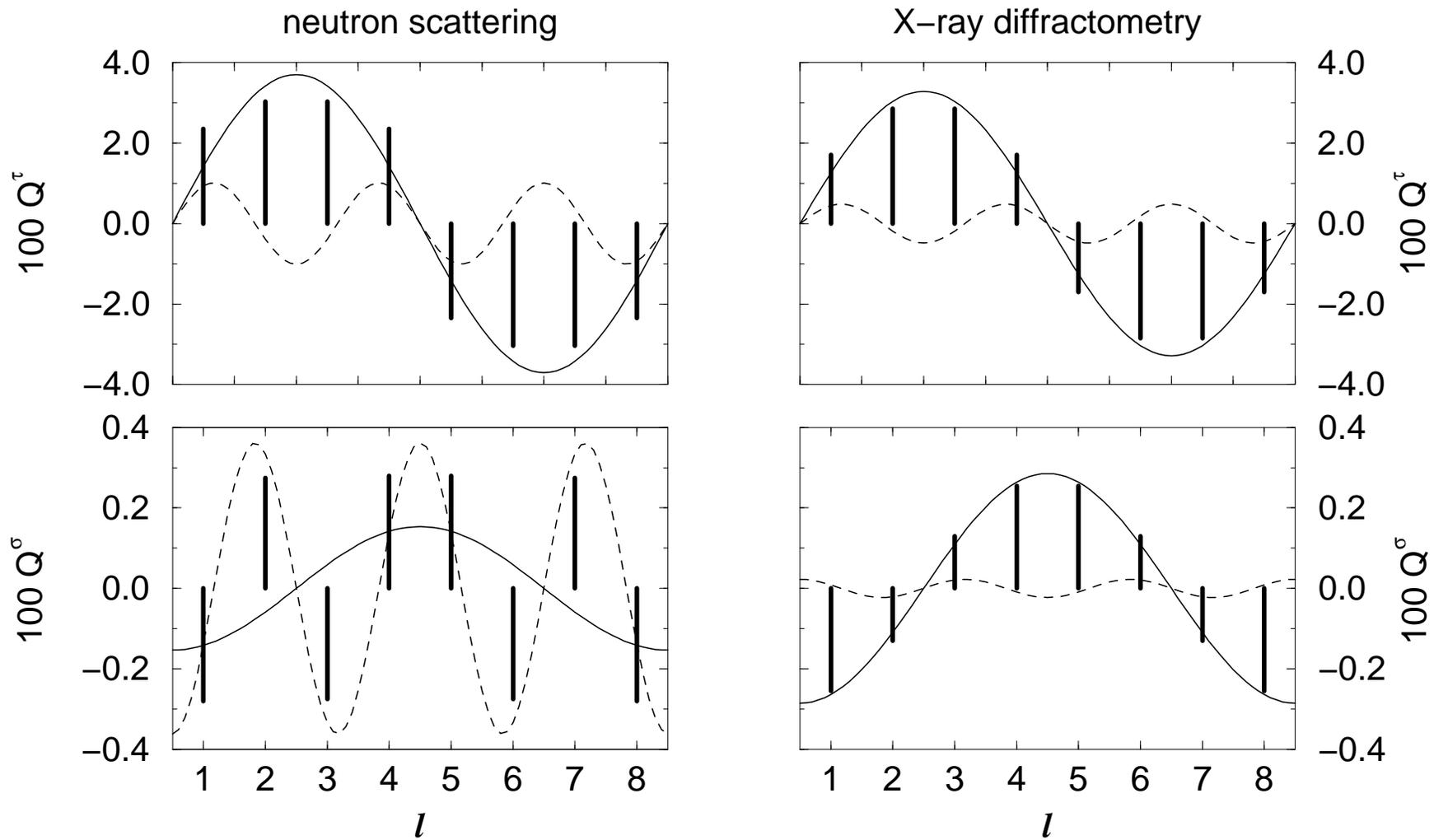
\end{document}